# Copper Large-scale Grain Growth by UV Nanosecond Pulsed Laser Annealing

Toshiyuki Tabata
*Laser Systems & Solutions of Europe (LASSE)*
Gennevilliers, France
toshiyuki.tabata@screen-lasse.com

Pierre-Edouard Raynal
*Laser Systems & Solutions of Europe (LASSE)*
Gennevilliers, France
pierre-edouard.raynal@screen-lasse.com

Fabien Rozé
*Laser Systems & Solutions of Europe (LASSE)*
Gennevilliers, France
fabien.roze@screen-lasse.com

Sébastien Halty
*Laser Systems & Solutions of Europe (LASSE)*
Gennevilliers, France
sebastien.halty@screen-lasse.com

Louis Thuries
*Laser Systems & Solutions of Europe (LASSE)*
Gennevilliers, France
louis.thuries@screen-lasse.com

Fuccio Cristiano
*LAAS-CNRS*
Toulouse, France
cfuccio@laas.fr

Emmanuel Scheid
*LAAS-CNRS*
Toulouse, France
scheid@laas.fr

*Fulvio Mazzamuto*
*Laser Systems & Solutions of Europe (LASSE)*
Gennevilliers, France
fulvio.mazzamuto@screen-lasse.com

*Abstract*—UV nanosecond pulsed laser annealing (UV NLA) enables both surface-localized heating and short timescale high temperature processing, which can be advantageous to reduce metal line resistance by enlarging metal grains in lines or in thin films, while maintaining the integrity and performance of surrounding structures. In this work UV NLA is applied on a typical Cu thin film, demonstrating a mean grain size of over 1 µm and 400 nm in a melt and sub-melt regime, respectively. Along with such grain enlargement, film resistivity is also reduced.

## I. Introduction

In advanced BEOL interconnects, reducing the trench geometry limits metal grain growth with the consequence of increasing electron scattering at grain boundaries. It results in an exponentially increasing line resistivity while scaling down the lines, and degradation of RC delay. In such scaling era, alternative metals such as Ru, Co, and Mo are introduced because of their potential benefits in line resistivity, which come from a complex combination of bulk resistivity, line width, mean-free-path of electrons, electro-migration reliability (i.e., melting point), integration compatibility, and especially the use of a specific set of barrier and liner [1-4]. However, even if the narrowest interconnects are formed with alternative metals, copper will not be completely replaced, and it will remain the reference to beat and the preferred candidate for larger interconnects. Thus, it is important to explore new paths to boost his performances and extend his utilization.

Extending Cu technology is still possible, especially by engineering the barrier/liner part [5-7]. On the other hand, nanosecond laser annealing (NLA) demonstrated a benefit on BEOL interconnects by enlarging the mean size of grains in both Cu [8,9] and Ru [10] lines. In fact, NLA allows to reach a much higher surface temperature than that of conventional BEOL limit (i.e., 400 °C for minutes), while conserving the functionality of surrounding devices thanks to its short timescale and shallow irradiation absorption. Such opportunity to reduce the interconnect resistance became particularly critical now, when the number of interconnect layers is continuously increasing [11].

In this paper we study high temperature processing realized by UV NLA to enable large-scale grain growth in thin films and lines (e.g., roughly 50-nm-thick). Specifically, we present the formation of large grains, which is, to our knowledge, a record for such a thin film (typically around 100 nm after annealing at 350 to 600 °C for minutes [12,13]). This opens a potential path to boost performances of future Cu interconnects.

## II. Experimental

A 50-nm-thick sputtered-Cu was deposited on 8 nm-thick Ta/5 nm-thick TaN/100 nm-thick $SiO_2$/Si without any capping layer on top. A UV NLA was performed at room temperature in air. Both laser fluence (*LF*) and process time (*t*) were varied to control the heat generated in the Cu film. The evolution of the material as a function of annealing condition was captured by in-plane XRD. Then, some selected conditions were analyzed by TEM and Electron Diffraction Mapping (EDM). Finally, the correlation between the film resistance and grain size was deduced.

## III. Results and Discussion

Figure 1 shows the XRD patterns taken at $LF_1$ for different $t$ ($t_1 < t_2 < t_3 < t_4 < t_5$), where the reference (i.e., non-annealed) data is also compared. In the as-deposited Cu film, the peaks of Cu (111), (200), and (220) planes are clearly observed, and a slight surface oxidation is also implied by the peak of $Cu_2O$ (111). For $t_1$ and $t_2$, the peaks of Cu (111) and (200) disappear, while those of Cu (220) grows. For $t_3$, $t_4$, and $t_5$, the disappeared peaks emerge again, while the peaks of Cu (220) show a significant drop of intensity. This suggests that the Cu film is melting (i.e., homogeneous nucleation) at the latter conditions ($t_3$, $t_4$, $t_5$), but not at the former ones ($t_1$, $t_2$). A similar XRD pattern evolution was observed at a smaller laser fluence ($LF_2$) as well (data not shown).

To get a more in-depth comparison of sub-melt and melt regimes, two UV NLA conditions (i.e., $t_2$ and $t_4$ at $LF_1$) were selected to pursue the analysis of microstructure in the Cu thin film. Figure 2 shows the plane-view TEM images of the non-annealed and annealed samples. In the as-deposited film, only small grains are observed everywhere. In the annealed films, a significant grain growth is observed. On these TEM specimens, electron diffraction patterns were obtained and fitted with those theoretically predicted for a face-centered-cubic (FCC) crystal lattice of Cu. Also, a mean grain size ($Av.$) was calculated as a weighted average of area ratio within each observed area, considering Σ3CSL facets as grain boundaries. In the as-deposited film (Fig. 3(a)(b)(c)), the small grains ($Av.$ 53.9 nm) are randomly distributed. After annealing at $LF_1$ for $t_2$ (Fig. 3(d)(e)(f)), a significant grain growth ($Av.$ 414 nm) is observed and interestingly the Cu (111) plane is preferably oriented on the top surface (i.e., ND). This is consistent with the fact that the (111) plane has the lowest surface energy in Cu FCC system [14]. In the in-plane directions (i.e., TD and RD), all the observed planes are on a typical tensile-strain-induced slip line (i.e., between [101] and [011] on an FCC-type stereogram) of Cu single crystal [15]. After annealing at $LF_1$ for $t_4$ (Fig. 3(g)(h)(i)), very large grains ($Av.$ 1000 nm) are randomly distributed. Then, the grain morphology seems drastically changed, and certain grains grow up to several μm scale in a preferred direction.

At the end, film resistivity was characterized for the same three samples. According to the literature [12], the Mayadas-Shatzkes (MS) model, which involves the scattering at grain boundaries, may give a better fit with our experimental data obtained in the 50-nm-thick Cu film, than the Fuchs-Sondheimer (FS) model. To start, the MS model was fitted with the film resistivity of the non-annealed sample by changing only the reflection coefficient, $R$. Then, $R = 0.45$ was obtained, which is close to the reported value ($R = 0.47$ [12]). The same $R$ was applied to the other samples. The effective thickness of the Cu and $Cu_2O$ layers before and after UV NLA were confirmed by cross-sectional TEM images. Figure 4 shows the fitting result as a function of the mean grain size ($Av.$). For the annealed samples, the film resistivity shows a deviation from the theoretical values. A possible origin of this deviation is associated to oxygen (O) contamination in our Cu thin films (before and) during the annealing. According to SIMS, the averaged O concentration level became higher when the mean grown grain size did larger (data not shown). In addition, in the melting regime, the Cu surface roughness may be significantly increased during UV NLA, resulting in more pronounced surface scattering of electrons compared to the non-annealed case. These two aspects will be improved in future works by covering the Cu surface with a capping layer and by better controlling the UV NLA conditions.

IV. CONCLUSION

We investigated the effect of UV NLA on a thin Cu film in order to control the grain growth and to find a path to mitigate the exponential growth of the metal resistance in advanced metal interconnects. In the sub-melt condition, the mean grain size ($Av.$) was increased up to 414 nm, almost 8-times-larger than that of the as-deposited film ($Av.$ 53.9 nm), with a controlled distribution of grain orientations. In the melting condition, the grain growth was extended further ($Av.$ 1000 nm), but the control of the grain orientation was not maintained. The observed grain growth led to a consistent reduction of the film resistivity. Although these results are promising, it is only the first fundamental study on thin films that motivates a further investigation. Particularly, the grain growth control needs to be confirmed in real interconnect structures. Also, it must be assured that the applied thermal budget does not degrade surrounding materials and structures.

ACKNOWLEDGMENT

This project has received funding from the ECSEL Joint Undertaking (JU) under grant agreement No 875999. The JU receives support from the European Union's Horizon 2020 research and innovation programme and Netherlands, Belgium, Germany, France, Austria, Hungary, United Kingdom, Romania, Israel.

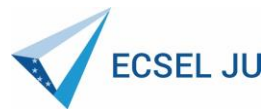
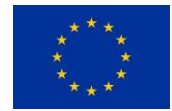

REFERENCES

[1] Z. Tokei, "Sub-5nm Interconnect Trends and Opportunities," IEDM 2017, short course.
[2] T. Nogami, "Overview of interconnect technology for 7nm node and beyond – New materials and technologies to extend Cu and to enable alternative conductors", EDTM 2019, pp. 38-40.
[3] C.-K. Hu *et al.*, "Electromigration and resistivity in on-chip Cu, Co and Ru damascene nanowires," IITC 2019.
[4] P. Bhosale *et al.*, "Co Reflow and Barrier Scaling Studies for Beyond 7 nm Node BEOL Applications," IITC 2019.
[5] C.-L. Lo *et al.*, "Replacing TaN/Ta Bilayer with 2D Layered TaS2 Converted from Ta for Interconnects at Sub-5 nm Technology Nodes," IITC 2019.
[6] K. Motoyama *et al.*, "EM enhancement of Cu interconnects with Ru liner for 7 nm node and beyond," IITC 2019.
[7] T. Nogami et al., "Advanced BEOL Interconnects," IITC 2020.
[8] R. T. P. Lee *et al.*, "Nanosecond Laser Anneal for BEOL Performance Boost in Advanced FinFETs," VLSI 2018, pp. 61-62.
[9] O. Gluschenkov *et al.*, "Laser Annealing in CMOS Manufacturing," ECS Trans., **85**(6), pp. 11-23 (2018).
[10] N. Jourdan *et al.*, "UV nanosecond laser annealing for Ru interconnects," IITC 2020.
[11] M J. Kobrinsky, "On-die interconnect challenges and opportunities for future technology nodes," VLSI 2020, short course.
[12] T. Sun *et al.*, "Surface and grain-boundary scattering in nanometric Cu films," Phys. Rev. B, **81**(15), 155454 (2010).
[13] J. S. Chawla et al., "Electron scattering at surfaces and grain boundaries in Cu thin films and wires," Phys. Rev. B, **84**(23), 235423 (2011).
[14] M. McLean and B. Gale, "Surface energy anisotropy by an improved thermal grooving technique," Philos. Mag., **20**(167), pp. 1033-1045 (1969).
[15] K. H. Kim and Y. M. Koo, "In-situ X-ray diffraction study of single-slipconditioned copper single crystals during uniaxial deformations," Philos. Mag. A, **81**(2), pp. 479-488 (2001).

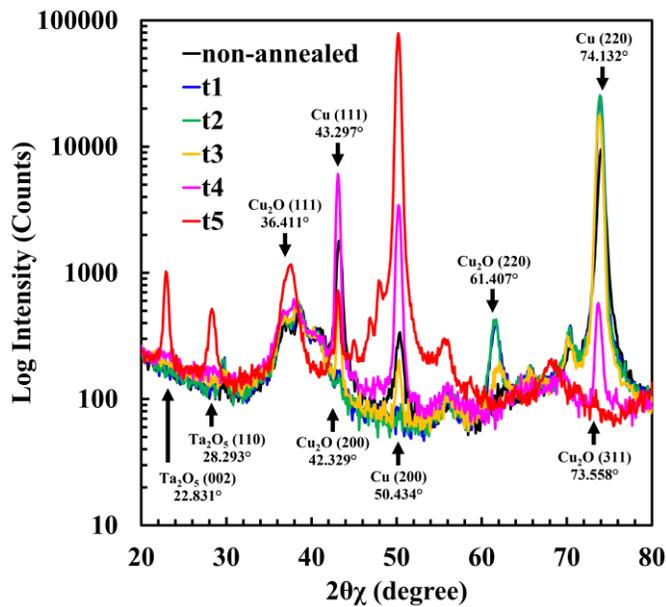

**Fig. 1.** XRD patterns obtained for the non-annealed and annealed Cu thin films. UV NLA was at $LF_1$ for different $t$ ($t_1 < t_2 < t_3 < t_4 < t_5$).

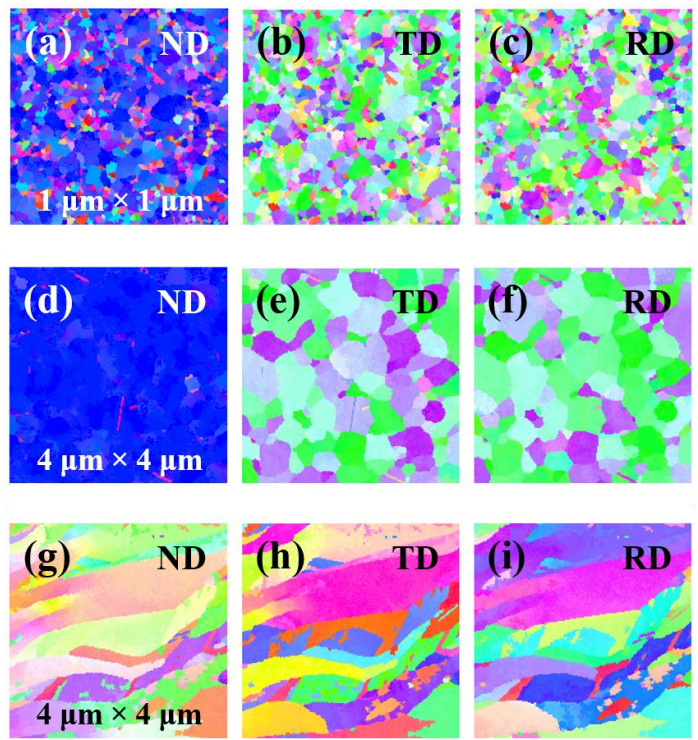

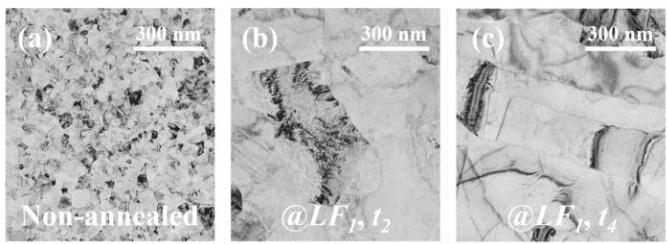

**Fig. 2.** Plain-view TEM images taken for the non-annealed and annealed Cu thin films. UV NLA was at $LF_1$ for $t_2$ and $t_4$.

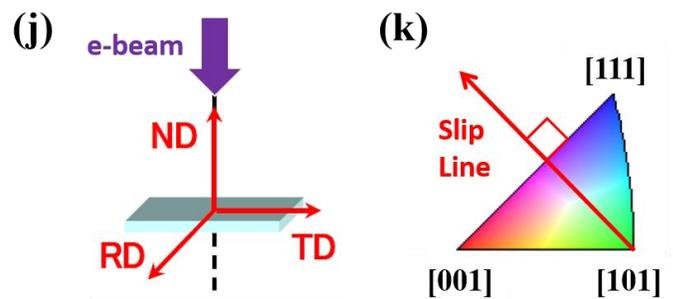

**Fig. 3.** EDM images taken for the non-annealed (i.e., (**a**), (**b**), and (**c**)) and the annealed Cu thin films. UV NLA was at $LF_1$ for $t_2$ ((**d**), (**e**), and (**f**)) or $t_4$ ((**g**), (**h**), and (**i**)). As depicted in (**j**), ND, TD, and RD stand for Normal Direction, Transverse Direction, and Reference Direction, respectively. Also, a standard triangle of grain orientations is also shown in (**k**).

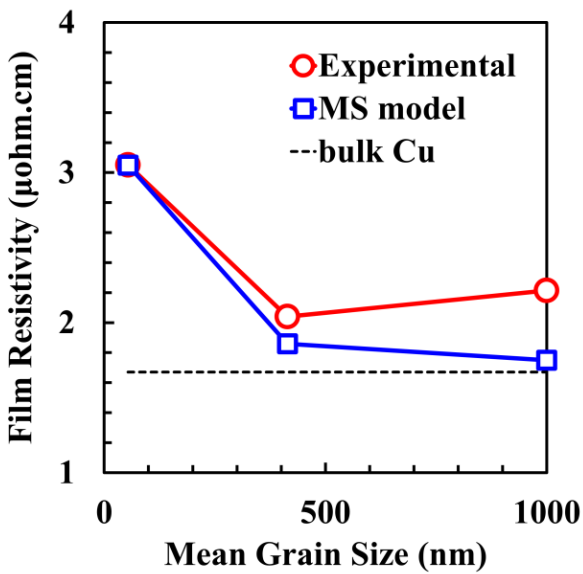

**Fig. 4.** Experimental and theoretical (based on the Mayadas-Shatzkes (MS) model) film resistivity of the non-annealed and annealed samples.